\documentclass[manuscript]{aastex}

\newcommand{\beq}{\begin{equation}}
\newcommand{\eeq}{\end{equation}}

\begin{document}

\title{WISE/NEOWISE Observations of the Jovian Trojans: Preliminary Results}

\shorttitle{WISE Observations of Jovian Trojans}
\shortauthors{Grav {\it et al.}}
\medskip

\author{T.~Grav} 
\affil{Department of Physics and Astronomy, Johns Hopkins University, Baltimore, MD21218, USA; tgrav@pha.jhu.edu}
\author{A.~K.~Mainzer, J.~Bauer\altaffilmark{1}, J.~Masiero} 
\affil{Jet Propulsion Laboratory, California Institute of Technology, Pasadena, CA 91109, USA}
\author{T.~Spahr}
\affil{Minor Planet Center, Harvard-Smithsonian Center for Astrophysics, Cambridge MA 02138, USA}
\author{R.~S.~McMillan}
\affil{Lunar and Planetary Laboratory, University of Arizona, Tucson, AZ 85721, USA}
\author{R.~Walker}
\affil{Monterey Institute for Research in Astronomy, Marina, CA 93933}
\author{R.~Cutri}
\affil{Infrared Processing and Analysis Center, California Institute of Technology, Pasadena, CA 91125, USA}
\author{E.~Wright} 
\affil{UCLA Astronomy, PO Box 91547, Los Angles, CA 90095, USA}
\author{E.~Blauvelt, E.~DeBaun, D.~Elsbury, T.~Gautier~IV, S.~Gomillion, E.~Hand}
\affil{Jet Propulsion Laboratory, California Institute of Technology, Pasadena, CA 91109, USA}
\author{A.~Wilkins\altaffilmark{2}}
\affil{Department of Astronomy, University of Maryland, College Park, MD 20742}

\altaffiltext{1}{Infrared Processing and Analysis Center, California Institute of Technology, Pasadena, CA 91125, USA}
\altaffiltext{2}{Jet Propulsion Laboratory, California Institute of Technology, Pasadena, CA 91109, USA}

%-------------------------------------------
\begin{abstract}
We present the preliminary analysis of over 1739 known and 349 candidate Jovian Trojans observed by the NEOWISE component of the Wide-field Infrared Survey Explorer (WISE). With this survey the available diameters, albedos and beaming parameters for the Jovian Trojans have been increased by more than an order of magnitude compared to previous surveys \citep{Tedesco.1992a,Tedesco.2002a,Fernandez.2003a,Fernandez.2009a,Ryan.2010a}. We find that the Jovian Trojan population is very homogenous for sizes larger than $\sim 10$km (close to the detection limit of WISE for these objects). The observed sample consists almost exclusively of low albedo objects, having a mean albedo value of $0.07\pm0.03$. The beaming parameter was also derived for a large fraction of the observed sample, and it is also very homogenous with an observed mean value of $0.88\pm0.13$. Preliminary debiasing of the survey shows our observed sample is consistent with the leading cloud containing more objects than the trailing cloud. We estimate the fraction to be $N(\mbox{leading})/N(\mbox{trailing}) \sim 1.4\pm0.2$, lower than the $1.6\pm0.1$ value derived by \citet{Szabo.2007a}.  
\end{abstract}
%-------------------------------------------
\section{Introduction}

The Jovian Trojan asteroids comprise two clouds around the L4 and L5 Lagrangian points in Jupiter's orbit. There are currently around 4800 known Jovian Trojans, with about half having multi opposition orbits. The population is believed to have a total number similar to that of the main-belt asteroids \citep[MBAs;][]{Tedesco.2005a,Yoshida.2005a,Jewitt.2000a}. The two clouds of Trojans librate around the L4 (leading cloud) and L5 (trailing cloud) Lagrange points with periods of the order of a few hundred years, and their orbital eccentricities ($<0.3$) and inclinations ($ < 40^\circ$) are similar to that of the MBAs. About three quarters of the Trojans that have been studied spectroscopically to date have feature-less (D-type) spectra \citep{Bus.2002a,Roig.2008a,Emery.2011a} that are found to have low optical albedo \citep{Tedesco.1989a,Sheppard.2003a,Fernandez.2003a,Fernandez.2009a}. The remaining fraction of the Trojans that have been spectroscopically characterized have P- or C-type spectra, mostly in the trailing swarm \citep{Fitzsimmons.1994a,Emery.2011a}. These spectral properties are similar to that of the cometary nuclei and are consistent with an origin in the outer Solar System. Studies of the size distributions support collisional grinding \citep{Jewitt.2000a}. 

The Jovian Trojans lie at the core of several of the most important aspects of planetary science, and there are several hypotheses have put forth to explain their origin: 1)  mutual collisions of planetesimals populating the region around Jupiter's orbit could have injected fragments into stable Trojan orbits \citep{Shoemaker.1989a}; 2) nebular gas drag could have produced drift of smaller planetesimals into the resonance gap, where they grew to present size through mutual collisions \citep{Yoder.1979a,Peale.1993a,Kary.1995a}; 3) they were formed simultaneously with Jupiter in the early phase of the solar nebula, where a growing Jupiter captured and stabilized the planetesimals near the L4 and L5 points \citep{Marzari.1998a, Marzari.1998b,Fleming.2000a}. \citet{Marzari.2002a} gives an excellent overview of the details of these different early proposed scenarios.     
      
It has also been suggested that, depending on the importance of gas drag during formation, the two clouds could have different dynamics, with the significant gas drag helping to stabilize orbits around the trailing cloud. Planetary migration, on the other hand, would destabilize the trailing cloud, causing it to evolve differently than the leading cloud \citep{Gomes.1998a}. More recently the so-called {\it Nice-model} suggested a more complex scenario: the current Trojan populations are objects that formed together with the Kuiper belt objects in a primordial disk ranging from roughly $\sim15-30$AU \citep{Morbidelli.2005a,Gomes.2005a,Tsiganis.2005a}. The Jovian Trojans were captured after the mutual 1:2 mean motion resonance crossing of Jupiter and Saturn during migration. This suggests the possibility that the physical and orbital properties of the leading and trailing clouds could be quite different. Such differences have yet to be found. It is, however, clear that the dynamical and physical distributions of the Trojan asteroids offer a critical window into differentiating between several models of Solar System dynamical evolution. 

It is important to note that there are severe observational biases in the sample of known Jovian Trojans. Jupiter and the Trojans take 12 years to complete a circuit around the ecliptic, and in the last decade the leading cloud has spent significantly more time in the northern hemisphere than the trailing cloud. During this time the number of known Trojans has increased ten-fold. With most of the optical large sky surveys located in the northern hemisphere, the leading cloud has seen significantly better coverage during this time. In addition, the trailing cloud has spent the last few years around the galactic center, an area that most surveys avoid due to the significant increase in star density that makes moving object identification correspondingly difficulty. While the number of MBAs and Trojans to a given size is similar, only about $1\%$ of the known asteroids are in the latter population. This ratio is a consequence of the larger distance that makes a trojan four magnitudes fainter than an MBA of similar size in the middle of the Main-Belt (this is not even accounting for differences in albedo that generally make the apparent magnitudes of the Trojans even fainter). 

In this paper we present the analysis of thermal measurements of more than 2000 known and candidate Jovian Trojans performed by the NEOWISE component of the {\it Wide-field Infrared Survey Explorer} \citep[WISE;][]{Wright.2010a,Mainzer.2011a}.  WISE, although a mission funded by the NASA's Astrophysical Division, is contributing significantly to the study of the Solar System, and has observed  more than 157,000 minor planets during its 1 year long survey. The large sample of Trojans with accurate thermal measurements will allow us to address the following questions: 1) What is the size distribution of the Jovian Trojans with diameters larger than WISE's detection limit of $\sim 5$km?; 2) Do the leading and trailing swarms have the same size and absolute number distribution above this detection limit?; 3) What is the albedo distribution of the WISE sample?; 4) Do the leading and trailing clouds have the same albedo distribution?

The WISE/NEOWISE observations are described in Section \ref{sec:observations}, and in Section \ref{sec:objsel} we describe the Trojan sample and how we select candidate Trojans from the sample of WISE/NEOWISE observations that do not have any optical follow-up. Section \ref{sec:thermalmodel} describes the thermal modeling in details. The analysis of the results of the thermal modeling is given in Section \ref{sec:results}.

\section{Observations}
\label{sec:observations}

\begin{figure}[t]
\begin{center}
\includegraphics[width=10cm]{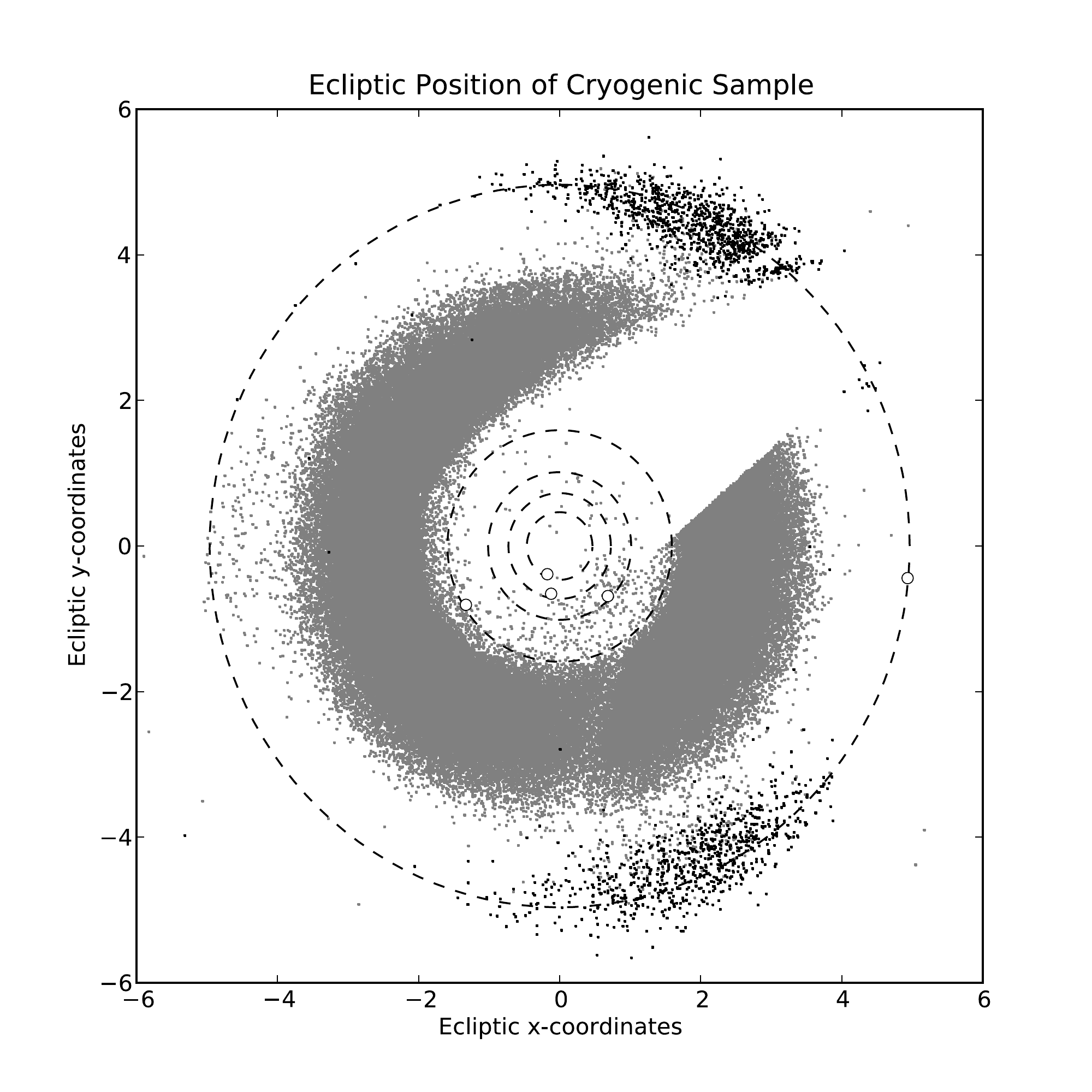}
\caption{The ecliptic x and y positions at 2010 August 5, of the 142,716 objects detected during the cryogenic part of the WISE survey that was assigned an orbit from the MPC (from 2010 January 14 to 2010 August 5). The two Jovian Trojans clouds are clearly seen $\sim60$ degrees leading and trailing the planet Jupiter. The orbits and positions of Mercury, Venus, Earth, Mars and Jupiter are also shown.}
\label{fig:ecliptic}
\end{center}
\end{figure}

WISE is a NASA Medium-class Explorer class mission designed to survey the entire sky in four infrared wavelengths, 3.4, 4.6, 12 and 22 $\mu$m \citep[denoted W1, W2, W3, and W4 respectively;][]{Wright.2010a,Lui.2008a,Mainzer.2005a}. The survey collected observations of over 157,000 asteroids, including Near-Earth Objects, Main Belt Asteroids, comets, Hildas, Jupiter Trojans, Centaurs and scattered disk objects \citep{Mainzer.2011a}. WISE has collected infrared measurements of nearly two orders of magnitude more asteroids than its predecessor, the {\it Infrared Astronomical Satellite} \citep[IRAS;][]{Tedesco.2002a, Matson.1989a}. The survey started on 2010 January 14 and the mission exhausted its secondary tank cryogen on 2010 August 5. The ecliptic x- and y-positions of the objects observed during the cryogenic part of the survey is shown in Figure \ref{fig:ecliptic}. Exhaustion of the primary cryogen tank occurred on 2010 September 29, but the survey was continued until 2011 February 1 as the NEOWISE Post-Cryogenic Mission using only bands W1 and W2, yielding a survey that observed the entire main-belt once. The WISE survey cadence resulted in most minor planets receiving on average of 10-12 observations over $\sim36$ hours \citep{Wright.2010a, Mainzer.2011a}. 

The WISE observations of the Trojans were retrieved by querying the Minor Planet Center (MPC) observation files to look for all instances of individual WISE detections of the desired objects that were reported using the WISE Moving Object Processing System \citep[WMOPS;]{Mainzer.2011a}. The resulting set of position/time pairs were used as the basis for a query of WISE source detections in individual exposures (known as ``Level 1b'' images) using the Infrared Science Archive (IRSA). To ensure that only observations of the moving objects were returned from the query, a search radius of $0.3"$ from the position listed in the MPC observation file was used. Furthermore, since WISE collected a single exposure every 11 seconds, the time of our observation was required to be within 4 seconds of the time specified by the MPC. Only observations with 0 or p in the artifact identification flag \textsc{cc\_flag} were used. A \textsc{cc\_flag} value of 0 indicates that no evidence of known artifacts was found at the position, while a \textsc{cc\_flag} of p indicates that an artifact may be present. We have found that observations with \textsc{cc\_flag} of p produce fluxes that are similar to non-flagged fluxes, resulting in recovery of $~20\%$ more observations. Some of the Trojans observed have W3 magnitudes smaller than 4, at which point the detector approached experimentally-derived saturation limits. In order to account for the inaccuracy of the point-spread-function fitting of these slightly saturated observations, the WISE W3 and W4 magnitude error bars were set to $0.2$ magnitudes \citep{Mainzer.2011b}.

In order to avoid having low-level noise detections and/or cosmic rays contaminating our thermal model fits we required each object to have at least three uncontaminated observations in a band. Any band that did not have at least 40\% of the observations of the band with the most numerous detections (W3 or W4 for the Trojans), even if it has 3 observations, was discarded. WMOPS was designed to reject inertially fixed objects such as stars and galaxies in bands W3 and W4. However, with stars having $\sim100$ times higher density in bands W1 and W2, it is more likely that asteroid detections in these bands are confused with inertial sources. We removed such confused asteroid detections by cross-correlating the asteroid detections with sources in the WISE atlas and daily co-added catalogs from IRSA. Objects within $6.5"$ (equivalent to the WISE beam size at bands W1, W2 and W3) of the asteroid position appearing in the co-added sources at least twice and in more than $30\%$ of the total number of coverages of a given area of sky were considered to be inertially fixed sources contaminating the asteroid positions, and these positions were removed from the thermal fitting. 

\section{Object Selection}
\label{sec:objsel}

\begin{figure}[t]
\begin{center}
\includegraphics[width=10cm]{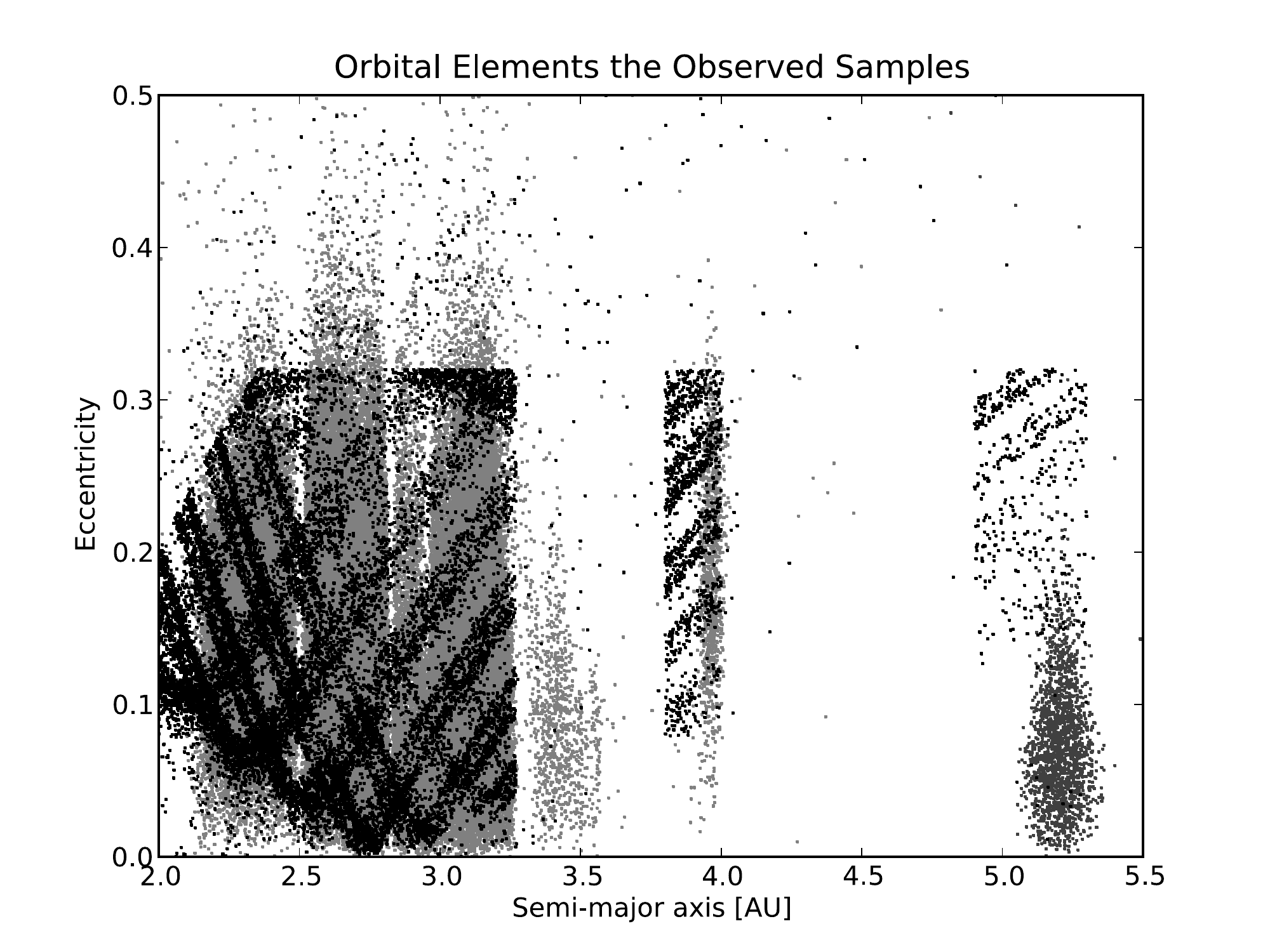}
\caption{The semi-major axes and eccentricities of the objects  detected with NEOWISE during the cryogenic part of the survey. The objects with length-of-arc longer than 18 days are given in grey. Objects with shorter length-of-arcs (SLA) are given in black and are mostly NEOWISE discoveries with no optical follow-up. The systematic pattern seen in the objects with short length-of-arcs (SLA) is an artifact of the MPC's discrete steps in assigning the elements of the highly uncertain orbits for these objects. }
\label{fig:aecc}
\end{center}
\end{figure}

\begin{figure}[t]
\begin{center}
\includegraphics[width=10cm]{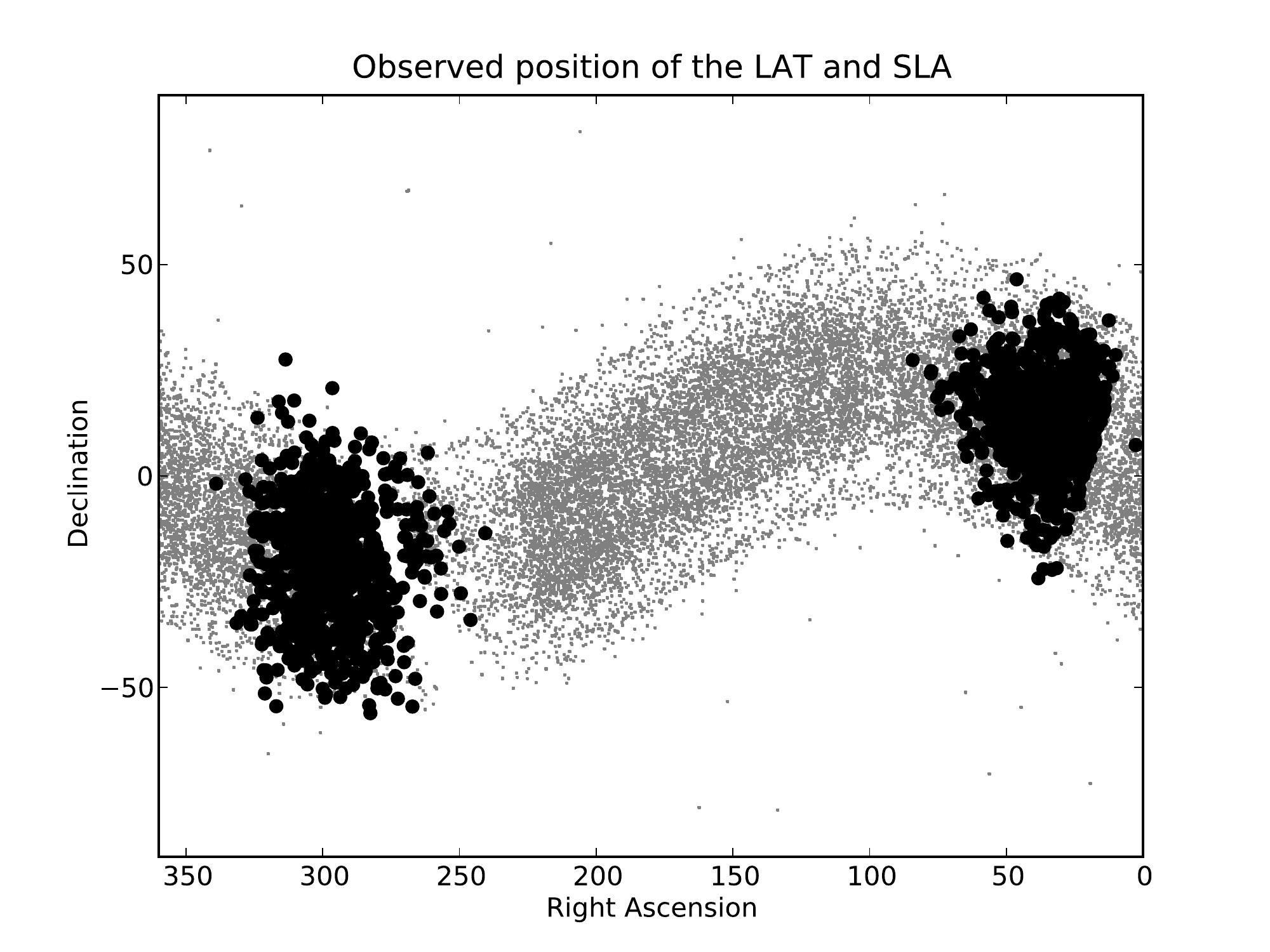}
\caption{The observed right ascension and declination of the short arc population (arc-lengths shorter than 18days) compared to the objects with long arc-lengths in the leading and trailing clouds. The effect of the galactic center can be seen clearly, with a dearth of detections near RA=269, DEC=-26 degrees. }
\label{fig:radec}
\end{center}
\end{figure}

As of 2011 April 5, there are currently 4846 known objects that the MPC has identified as Jovian Trojans. Some of these have observational arc lengths that makes their identification tenuous at best. During the cryogenic part of its survey NEOWISE observed 1751 objects that have orbits with semi-major axis between 5.0. and 5.4 AU with arcs  long enough to securely identify them as Trojans (here set somewhat arbitrarily to 18 days, with 981 of these in the leading and 770 in the trailing cloud. In the rest of the paper we will call this sample the long-arc Trojans (LAT).  

\begin{figure}[htbp]
\begin{center}
\includegraphics[width=10cm]{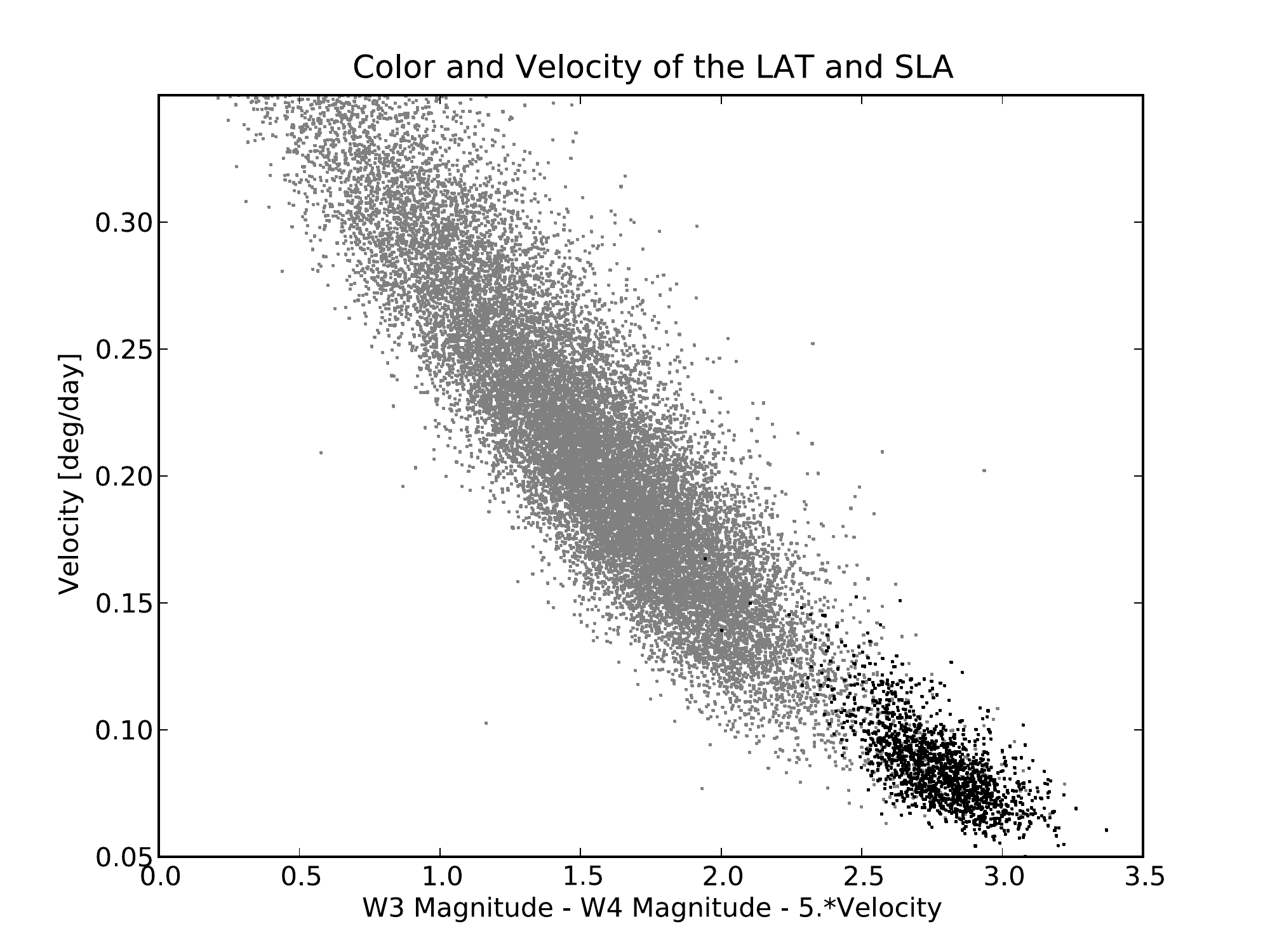}
\caption{The W3 and W4 magnitude color is plotted versus the sky-plane velocity of the SLA (grey) and LAT (black) samples. This distribution shows that we can do a cut at sky-plane velocity $V< 0.18$ and $W3-W4-5*V > 2.25$, W3 and W4 are the magnitudes in the two bands, to derive the SAT sample.}
\label{fig:velw34}
\end{center}
\end{figure}

In addition, NEOWISE detected 20,685 objects with length-of-arcs of less than 18 days (here donated the SLA sample). Most of these short-arc objects  are NEOWISE discoveries and their orbits remain highly uncertain with clear non-natural feature seen in their orbital parameter distributions caused by the discrete steps the MPC uses to assign their orbital elements (see Figure \ref{fig:aecc}). However, there exist a few ways to use the observed quantities to remove objects that cannot be associated with the Jovian Trojan clouds. Figure \ref{fig:radec} shows the observed right ascension and declination of this short length-of-arcs sample and the LAT. Imposing a set of right ascension criteria reduces the number of significantly. Furthermore, due to their heliocentric distance any Jovian Trojan will be moving at very low sky-plane velocities. Using the long-arc members of the cloud we set the velocity limit to $< 0.18$ degrees per day. This criterion reduces the short-arc sample to 1722 objects (see Figure \ref{fig:velw34}). We then use the thermal color measurements to further remove objects that are inconsistent with the LAT sample. We use $W3 - W4 - 5v > 2.25$, where $v$ is the on-sky velocity in degrees per day. It is important to note that these criteria could remove potential trojans that have physical properties very different than that of the LAT sample. However, it seems unlikely that there would be a large number of such objects that would affect our ability to debias the survey.  These criteria yield a sample of 349 objects, 208 in the leading cloud and 141 in the trailing cloud, and in the rest of the paper we call this sample the short-arc Trojans (SAT). It is expected that this SAT sample will receive additional follow-up as the major surveys observe the Jovian Trojan clouds in the future, allowing the MPC to link our observations to those of the optical surveys. It should also be noted that the SAT are only candidate objects. Many of the longer length-of-arc Hildas and even some MBAs have velocities and colors consistent with the LAT sample and so a fair fraction of the SAT sample is expected to be from these populations rather than the Jovian Trojans. 

NEOWISE has reported 16,556 observations of the 1751 Trojans with well defined orbits. We extracted the observations from the archive using the method described in \citet{Mainzer.2011b} and received 16,551 data lines. There are 2949 observations in the MPC observation catalog of the 349 possible Jovian Trojans. We extracted the observations from the archive and found 2946 corresponding data lines.

\section{Preliminary Thermal Modeling}
\label{sec:thermalmodel}

Preliminary thermal models for each of the Trojans observed by WMOPS during the cryogenic portion of the survey and using the First-Pass Data Processing Pipeline (version 3.5) described above has been computed (these thermal models will be recomputed when the final data processing is completed some time during the fall of 2011). As described in \citet{Mainzer.2011b} the spherical near-Earth asteroid thermal model \citep[NEATM;][]{Harris.1998a} was used. The NEATM  uses the so-called beaming parameter $\eta$ to account for cases intermediate between zero thermal inertia \citep[the Standard Thermal Model, or STM;][]{Lebofsky.1978a} and high thermal inertia \citep[the Fast Rotating Model, or FRM;][]{Veeder.1989a,Lebofsky.1989a}. In the STM, $\eta$ is set to 0.756 to match the occultation diameters of Ceres and Pallas, while in the FRM, $\eta$ is equal to $\pi$. In the NEATM $\eta$ is a free parameter than can be fitted if two or more thermal bands are available, or using a single thermal band if {\it a priori} information of diameter and albedo is available from space craft or occultation observations. The effects of rotational variability are discussed in more detail below.

For each object a spherical surface was approximated using a set of triangular facets \citep[c.f.][]{Kaasalainen.2004a}. While Trojans may be significantly non-spherical, the WISE observations generally consist of $~8-10$ observations uniformly distributed over $\sim36$h  for each object, such that rotational variation in generally is averaged out. Caution needs to be exercised when interpreting the meaning of an effective diameter in cases of objects with higher rotational amplitudes \citep{Wright.2007a}. 

Thermal models were computed for each individual WISE measurement, ensuring that the correct Sun-observer-object geometry were used. The temperature of each facet was computed and color corrections were applied to each facet based on \citet{Mainzer.2011b}. Nightside facets were assumed to contribute no thermal flux. Adjustments of the W3 effective wavelength blue-ward by $4\%$ from $11.5608\mu$m to $11.0984\mu$m, and the W4 effective wavelength red-ward by $2.5\%$ from $22.0883\mu$m to $22.6405\mu$m, were used. In addition the $-8\%$ and $+4\%$ offsets to the W3 and W4 magnitude zeropoints (respectively) due to the red-blue calibrator discrepancy were also used \citep{Wright.2010a,Mainzer.2011b}. In general, orbital elements and absolute magnitudes were taken from the MPC catalogs, and we assumed the absolute magnitude $H$ to have an error equal to $0.3$ magnitudes. Emissivity, $\epsilon$, was assumed to be 0.9 for all wavelengths \citep[c.f. ][]{Harris.2009a}, and the slope parameter, G, in the magnitude-phase relationship \citep{Bowell.1989a} was set to 0.15 unless an improved value exists in the MPC catalogs. 

\begin{figure}[t]
\begin{center}
\includegraphics[width=10cm]{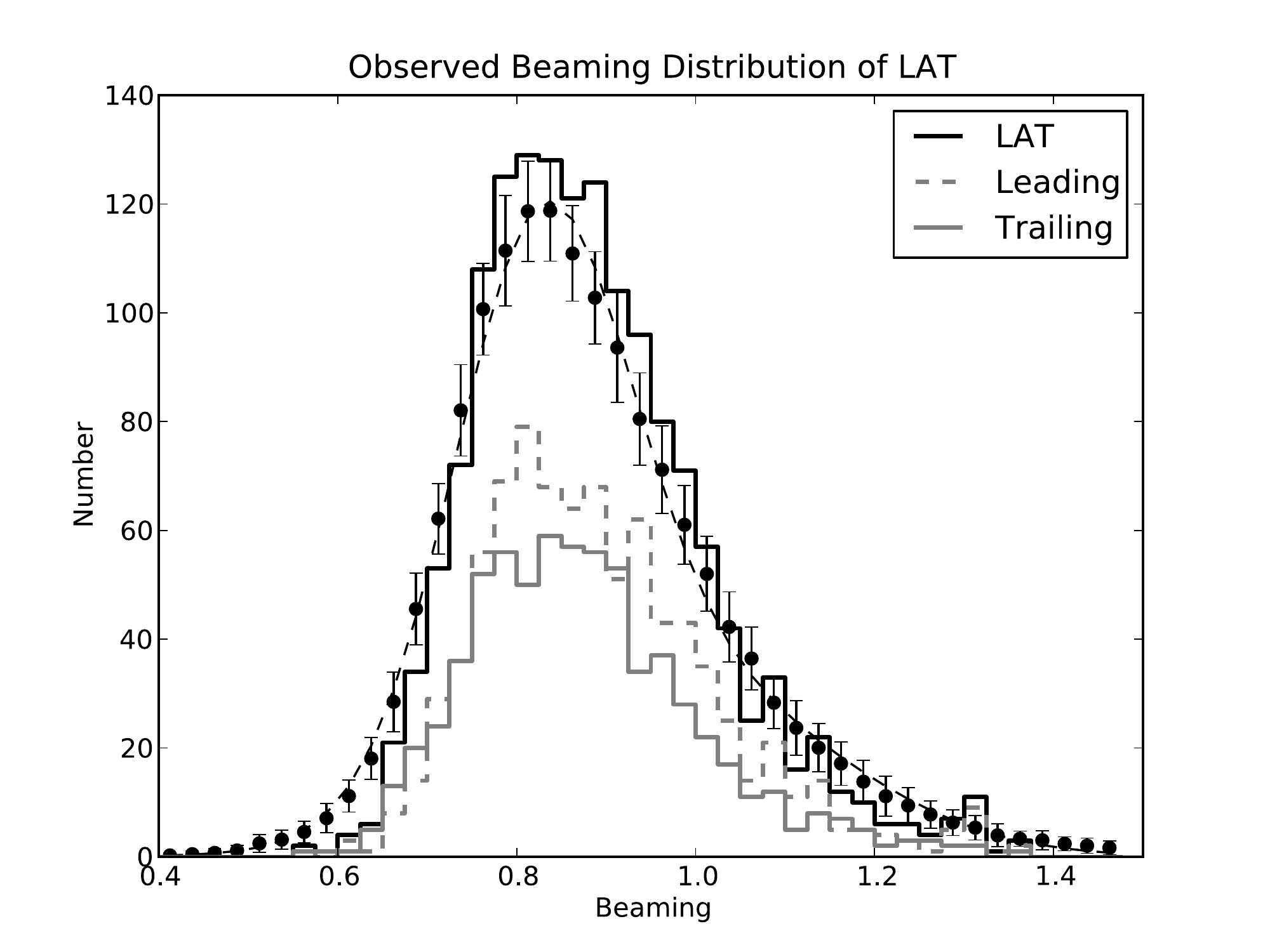}
\caption{The beaming distribution of  the  long-arc Trojans (LAT) objects for which beaming was derived. The leading and trailing clouds are shown in grey dashed and grey solid, respectively. The solid points are derived by generating 100 Monte-Carlo (MC) trials to yield different sample beaming distributions by varying the individual beaming values by their errors. For each trial distribution the histogram was generated, and the set of 100 histograms was used to derive the mean value and standard deviation in each bin. A double Gaussian distribution was then fitted to these MC mean values and associated standard deviations, plotted as a dashed black line.}
\label{fig:etahist}
\end{center}
\end{figure}

For Jovian Trojans with measurements in both W3 and W4, the beaming parameter $\eta$ was determined using a least square minimization (but was constrained to be less than the upper bound set by the FRM case, $\pi$).  Figure \ref{fig:etahist} shows the fitted $\eta$ value histogram for the objects with both thermal bands, along with the best fitting double Gaussian distribution. The median value of the 1534 objects in the LAT with fitted $\eta$ is $0.88\pm0.13$, while the weighted mean value is $0.84\pm 0.11$. The best-fit double Gaussian shown in Figure \ref{fig:etahist} has a mean value of $0.84\pm 0.10$ and $0.97\pm0.18$ with the lower mean Gaussian having a peak $\sim3$ times higher than the higher mean. For the 216 objects in the LAT with only one thermal measurement, the beaming value cannot be fitted, and we assumed a value of $0.87\pm0.13$. 

\begin{figure}[t]
\begin{center}
\includegraphics[width=10cm]{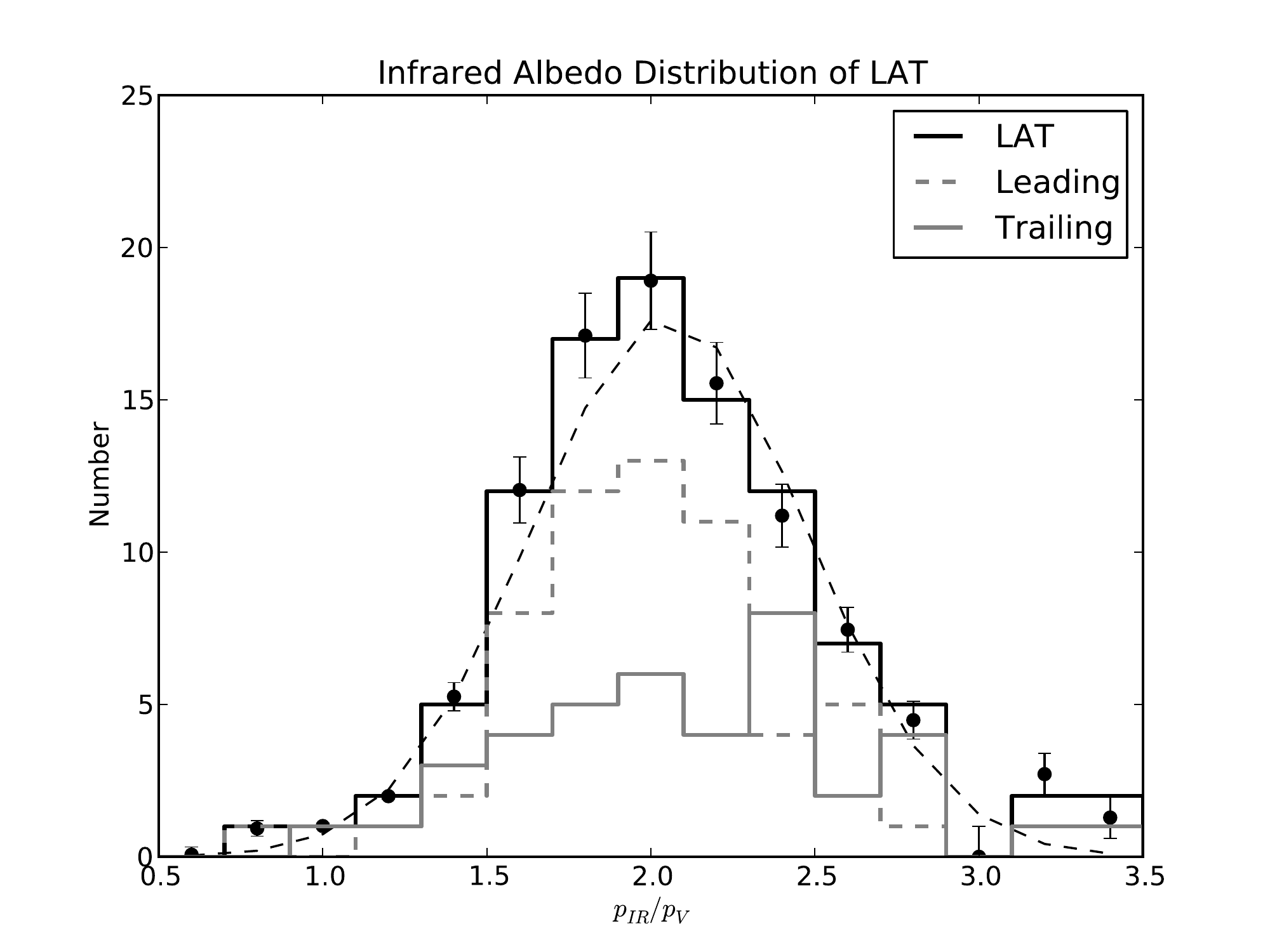}
\caption{The IR albedo/Visible albedo distribution of  the  100 long-arc Trojans (LAT) objects for which $p_{IR}$ was derived. The leading and trailing clouds are shown in dashed and solid grey, respectively. The solid points are derived by generating 100 MC trials to yield different sample beaming distributions by varying the individual beaming values by their errors. For each trial distribution the histogram was generated, and the set of 100 histograms was used to derive the mean value and standard deviation in each bin. A single Gaussian distribution was then fitted to these MC mean values and associated standard deviations, plotted as a dashed black line.}
\label{fig:irfachist}
\end{center}
\end{figure}

For the Jovian Trojans, bands W1 and W2 are generally dominated by reflected light. The flux from reflected sunlight was computed for each WISE band as described in \citet{Mainzer.2011b} using the IAU phase curve correction \citep{Bowell.1989a}. Those facets that were illuminated by reflected light and observable by WISE were corrected using color corrections appropriate for a G2 V star \citep{Wright.2010a}. In order to compute the fraction of total luminosity due to reflected light, the albedo in W1 and W2, dubbed $p_{IR}$ was introduced (we assume that $p_{IR}$ is the same for both bands). The geometric albedo $p_V$ is defined as the ratio of brightness of an object observed at zero phase angle to that of a perfectly diffusing Lambertian disk of the same radius located at the same distance. Related to the visible geometric albedo, is the Bond albedo, A, given by $A \approx A_V = qp_V$, where the phase integral $q$ is given by $0.290+0.684G$ \citep{Bowell.1989a}. The albedo in W1 and W2, $p_{IR}$, is assumed to obey the same relationship, although it is possible that it varies with wavelength, so what we denote here as $p_{IR}$ for convenience may not be exactly analogous to $p_V$ . The resulting distribution is shown in Figure \ref{fig:irfachist} and the mean of the $p_{IR}/p_V$ value for the 100 objects for which we were able to fit both $p_{IR}$ and $p_V$ is $2.0\pm0.5$.

The error in the fits of diameter ($D$), albedo ($p_V$), W1/W2 albedo ($p_{IR}$) and beaming ($\eta$) were determined for each object by running 50 Monte Carlo (MC) trials that varied the object's absolute magnitude ($H$) values by the errors described above and the WISE magnitudes by their error bars using Gaussian probability distributions. The minimum magnitude error for all WISE measurements was set to $0.03$ magnitudes, in line with the repeatability given in \citet{Wright.2010a}. If the source was brighter than the saturation limits of $6$, $6$, $4$ or $3$ in bands 1 to 4, respectively, the magnitude error of that band was set to 0.2 magnitude to reflect the tests performed in \citet{Mainzer.2011b} using calibration objects with known diameters. For objects with fixed $\eta$, errors on derived parameters were computing by varying $\eta$ by a gaussian centered on $0.875$ with a full width half max (fwhm) of 0.13. For objects which the W1/W2 albedo, $p_{IR}$ could not be fit, the MC trials varied $p_{IR}/p_V$ using a gaussian with center at $2.0$ and a fwhm of $0.5$. 

As described in \citet{Mainzer.2011b} the minimum diameter error that can be achieved using WISE observations is $\sim 10\%$, while the minimum albedo error is $\sim 20\%$ of the stated value, where two thermal bands are available and $\eta$ can be fitted.
Since a significant fraction of the Trojan population has been found to have non-spherical shapes \citep{Hartmann.1988a,Binzel.1992a, Mottola.2011a} some care has to be taken in interpreting the derived values presented herein. All diameters given are considered effective diameters, where the assumed sphere has a volume close to that of the actual body observed. Tests using a variety of synthetic triaxial ellipsoidal bodies with different sizes, elongations and pole orientations show that even for objects with significant rotational lightcurves, $\sim1$ magnitude, the effective diameter derived is generally found to have a 1 sigma error bar of $\sim20\%$ when compared to the spherical-equivalent diameter of the highly elongated ellipsoidal test bodies. This results holds as long as the rotational period is not significantly lower than our sample rate of $\sim3$ hours or significantly longer than the average coverage of an object of $\sim36$ hours. For the shorter rotational periods, the error quoted above is still generally valid, but for the longer rotational periods the derived effective diameters can be anywhere from the highest to the lowest extent of the axes depending on what part of the rotational lightcurve our sample covered. Identifying objects with short or long rotational periods will have to be done in conjunction with existing or new optical lightcurve data, as WISE/NEOWISE in general does not sample often enough to determine periods shorter than the spacecraft's orbital period and does not cover time spans long enough to sample enough of the lightcurve of the objects with periods longer than $\sim 36$ hours. A more comprehensive study of the rotational lightcurves by combining existing and new optical data with the thermal data collected from WISE/NEOWISE and the influence of these lightcurves on the individual objects' fits will be covered in a future paper. However, the results presented herein are a statistically valid sample of effective diameters and albedos for the Jovian Trojan population. 

Table \ref{ex:etable} shows some examples of the results of the thermal model fits and a full electronic version of the table for the 1739 Jovian Trojans detected by WMOPS using the First-Pass processing pipeline during the cryogenic WISE/NEOWISE mission and that have the necessary filtered observations is available at the journal website. 

\section{Results}
\label{sec:results}

\begin{table}[htdp]
\caption{Example of Electronic Table of the Thermal Model Fits }
\begin{center}
\begin{tabular}{lcccccccc}
Object & H & G & D & $p_V$ & $\eta$ & $p_{IR}$ & \# obs & mJD \\
\hline 
00588 &  8.67 & 0.15 &  $160.6\pm11.9$ & $0.023\pm0.006$ & $1.02\pm0.09$ & $0.076\pm0.010$ &  9  9  9  9 & 55205.5 \\
00617 &  8.19 & 0.15 &  $185.1\pm13.1$ & $0.027\pm0.006$ & $0.90\pm0.08$ & $0.057\pm0.007$ & 10 10 11 11 & 55281.5 \\
00624 &  7.20 & 0.15 &  $163.9\pm 7.2$ & $0.087\pm0.016$ & $0.96\pm0.05$ & $0.107\pm0.012$ & 11 11 11 11 & 55228.2 \\
00659 &  8.99 & 0.15 &  $122.4\pm10.7$ & $0.030\pm0.007$ & $0.93\pm0.10$ & $0.059\pm0.008$ &  8 11 11 11 & 55215.4 \\
00884 &  8.81 & 0.15 &  $116.9\pm7.6$ & $0.039\pm0.012$ & $1.01\pm0.08$ & $0.092\pm0.012$ &  9 10 10 10 & 55317.7 \\
00911 &  7.89 & 0.15 &  $143.8\pm4.5$ & $0.060\pm0.012$ & $1.07\pm0.04$ & $0.094\pm0.008$ & 21 21 21 21 & 55221.2 \\
01172 &  8.33 & 0.15 &  $138.7\pm3.9$ & $0.043\pm0.009$ & $1.03\pm0.03$ & $0.088\pm0.009$ &  9 10 10 10 & 55315.7 \\
01173 &  8.89 & 0.15 &  $114.4\pm8.0$ & $0.038\pm0.009$ & $0.88\pm0.12$ & $0.075\pm0.019$ &  0  0  0  7 & 55277.6 \\
01208 &  8.99 & 0.15 &  $134.1\pm6.1$ & $0.025\pm0.006$ & $0.86\pm0.05$ & $0.059\pm0.007$ & 11 12 12 12 & 55303.5 \\
01404 &  9.30 & 0.15 &  $88.4\pm3.0$ & $0.043\pm0.008$ & $0.93\pm0.04$ & $0.065\pm0.008$ & 12 12 12 12 & 55209.7 \\
01437 &  8.30 & 0.15 &  $128.1\pm9.8$ & $0.052\pm0.012$ & $1.00\pm0.10$ & $0.103\pm 0.024$ &  0  0 10 10 & 55244.1 \\
\hline
\end{tabular}
\end{center}
\label{ex:etable}
\end{table}%

\begin{figure}[t]
\begin{center}
\includegraphics[width=10cm]{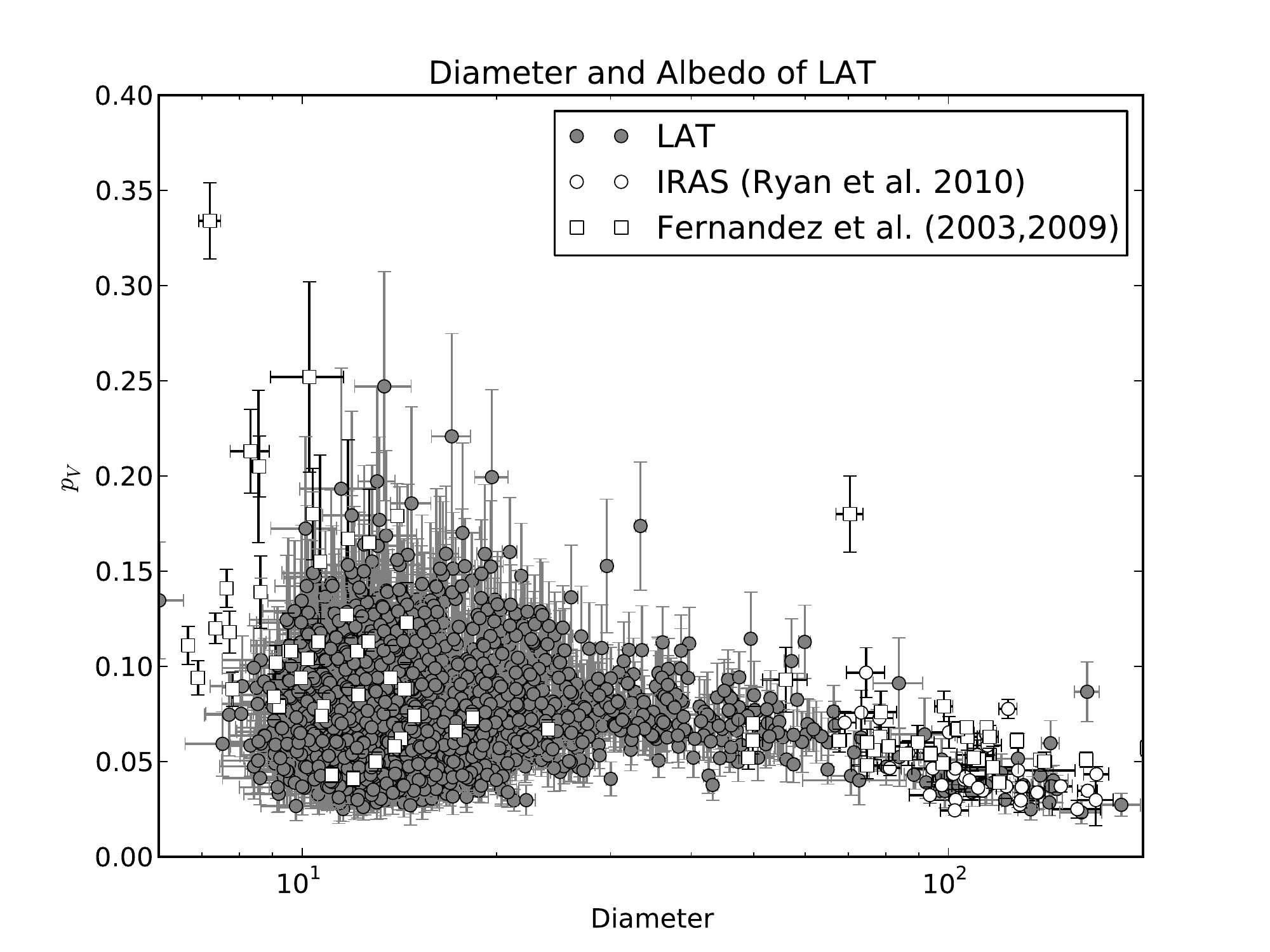}
\caption{The 1739 long-arc Trojans (LAT) objects for which diameter and albedo was derived.  Diameter and albedos derived by \citet{Fernandez.2003a,Fernandez.2009a} and \citet{Ryan.2010a} are shown for comparison.}
\label{fig:DpV}
\end{center}
\end{figure}

Thermal models were derived for 1739 Jovian Trojans from the LAT sample (for 9 objects there were not enough uncontaminated detections to derive thermal fits), with 985 objects in the leading cloud and 754 objects in the trailing cloud. The diameter versus albedo distribution is shown in Figure \ref{fig:DpV} . The trojans are compared to the IRAS result \citep{Ryan.2010a} and ground based and Spitzer results from \cite{Fernandez.2003a,Fernandez.2009a}, and are in very good agreement. Our sample does not have the higher albedo seen in some of the smaller objects in the \citet{Fernandez.2009a} sample. 

\begin{figure}[t]
\begin{center}
\includegraphics[width=10cm]{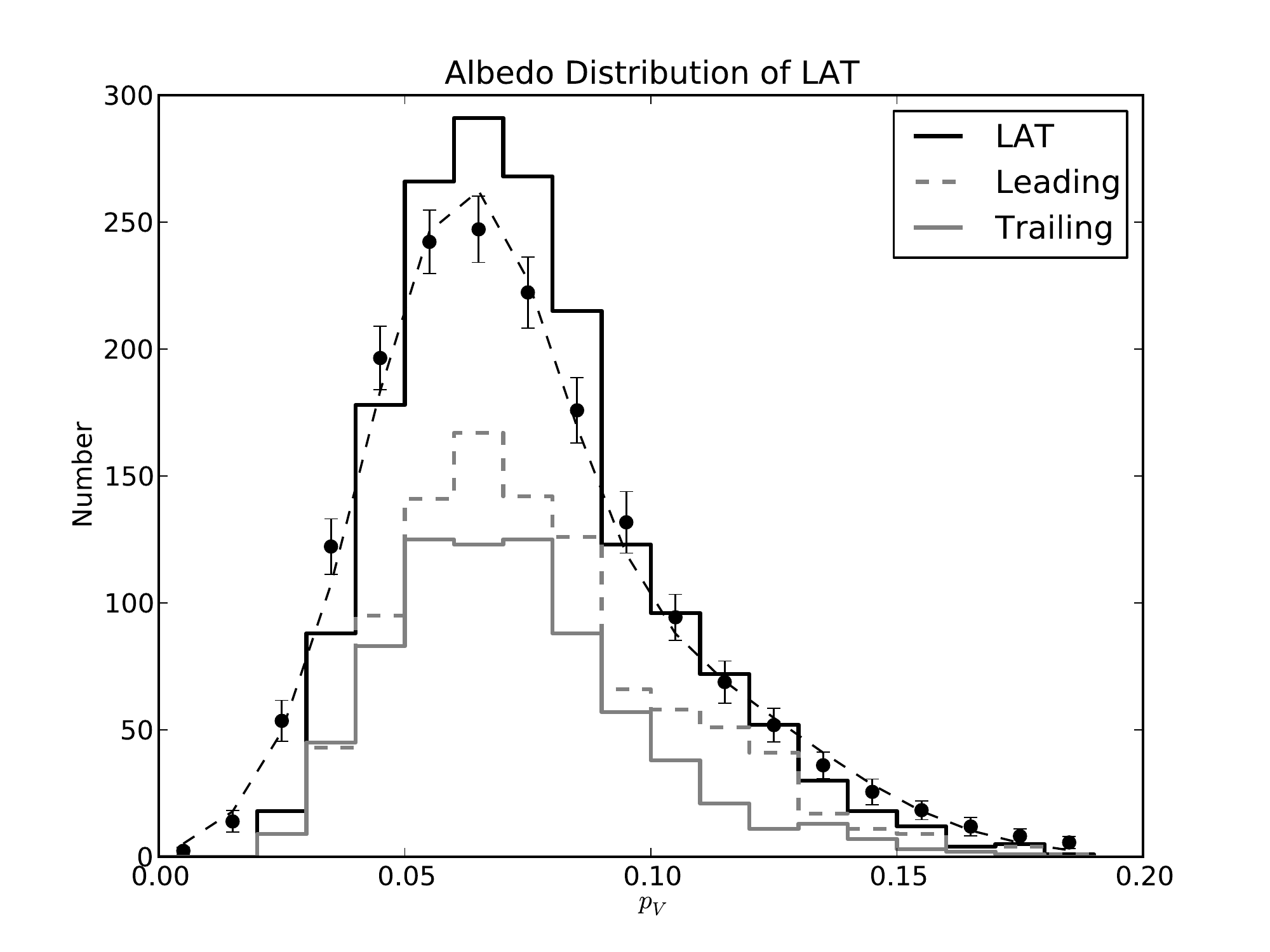}
\caption{The albedo distribution of  the long-arc Trojans (LAT) objects for which albedo was derived. The leading and trailing clouds are shown in dashed and solid grey, respectively.}
\label{fig:pVhist}
\end{center}
\end{figure}

The best-fit albedo distribution of the 1739 objects in the LAT sample is given in Figure \ref{fig:pVhist}. The mean value for the entire sample $0.07 \pm 0.03$, with both the leading and trailing cloud having the same value. We attempt to fit the sample to a double Gaussian sample. First we used 50 Monte-Carlo trials to derive different sample albedo distributions by varying the individual albedos by its error. For each trial distribution the histogram was generated, and the set of 50 histograms was used to derive the mean value and standard deviation in each bin. The results are shown in Figure \ref{fig:pVhist} as the points with error bars. We then attempted to fit a double Gaussian distribution to these mean values. We find that the LAT sample is best fit with a low albedo Gaussian with mean value of $0.06$ and a fwhm of $0.02$ and a higher albedo Gaussian with mean value of $0.10$ and fwhm of $0.3$. The peak amplitude of the two Gaussians has a ratio of $\sim 3.$ in favor of the low albedo distribution.

The mean value is slightly higher than the historically canonical value of $0.040\pm0.005$ \citep{Tedesco.1989a,Jewitt.2000a}. The albedos of the larger objects are, however, consistent with the albedo derived by other authors \citep{Ryan.2010a,Fernandez.2009a} The albedos of the small objects are also consistent with that of \citet{Fernandez.2003a}, although that project found a few small objects with albedos that are on the high end of our sample. 

The homogeneously low albedos strengthen the hypothesis that the Jovian Trojans consist of the low albedo C-, D- and P-type asteroids \citep{Tedesco.1989a, Sheppard.2003a,Fitzsimmons.1994a,Emery.2011a}. No difference is, however, found among the visible albedo distributions of the leading and trailing clouds that indicate any differences in the taxonomic distribution between the two clouds as suggested by \citet{Fitzsimmons.1994a} and \citet{Emery.2011a}. It could simply be that there is no good way to distinguish between these low albedo taxanomic types based on visible albedo alone \citep{Mainzer.2011d}. We will, however, study the correlation of albedo with taxonomic class, spectral slope and broadband colors in a future paper. 

\begin{figure}[t]
\begin{center}
\includegraphics[width=10cm]{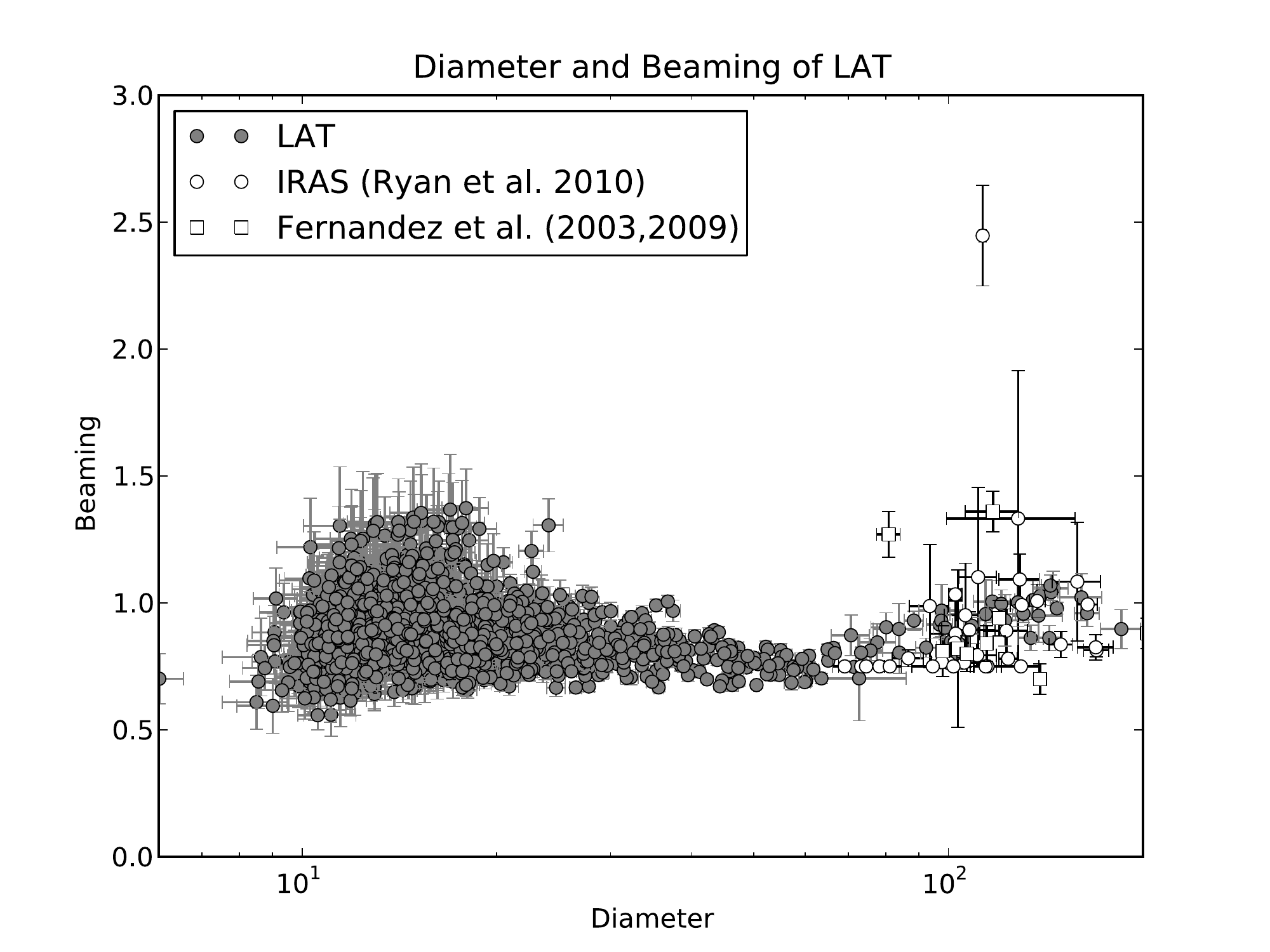}
\caption{The  long-arc Trojans (LAT) objects for which diameter and beaming value were derived. Diameter and beaming value derived by  \citet{Fernandez.2003a} and \citet{Ryan.2010a} are shown for comparison.}
\label{fig:Deta}
\end{center}
\end{figure}

Two thermal bands were available for 1523 objects (831 in the leading and 692 in the trailing cloud), allowing us to derive beaming values. The results are shown in Figure \ref{fig:Deta} and comparing the beaming value for these objects with those derived by \citet{Fernandez.2003a} and \citet{Ryan.2010a} show that our values are generally consistent with those found by these authors. The distribution of beaming values is shown in Figure \ref{fig:etahist} and is very similar for the two clouds. The mean beaming value is $0.89$ for the leading cloud, $0.87$ for the trailing cloud and $0.88$ for the full sample. The beaming values for the population is very well defined with standard deviation of $\sim0.13$ in both clouds and the full sample. This low dispersion points to the fact that the thermal properties of the Jovian Trojan LAT sample on a whole are very homogeneous compared to the Near-Earth Objects \citep{Mainzer.2011c} and Main-Belt Asteroids \citep{Masiero.2011a}. 

\begin{figure}[t]
\begin{center}
\includegraphics[width=10cm]{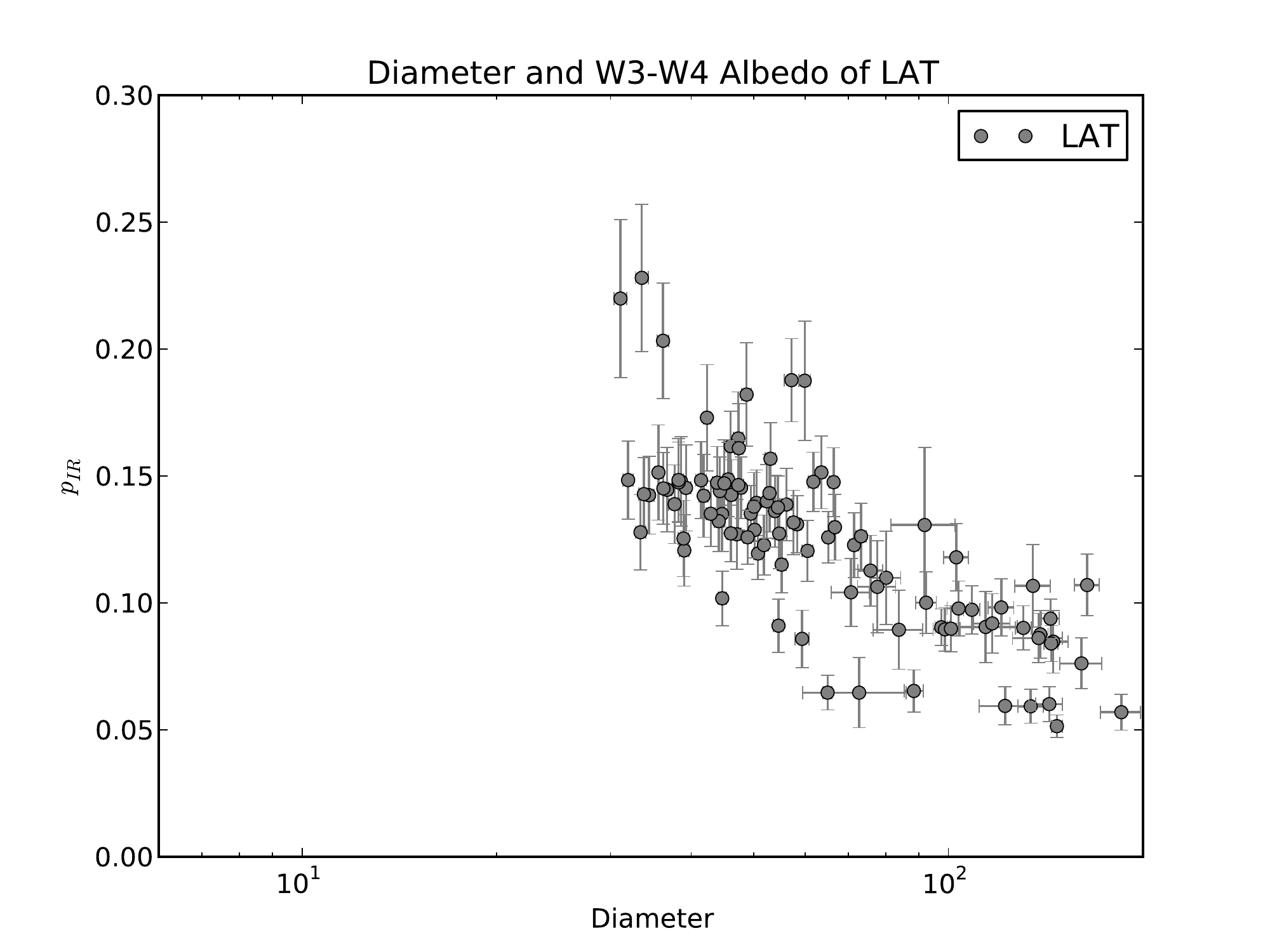}
\caption{The 100 long length-of-arc Trojans (LAT) objects for which diameter and infrared albedo were derived.}
\label{fig:DpIR}
\end{center}
\end{figure}

\begin{figure}[t]
\begin{center}
\includegraphics[width=10cm]{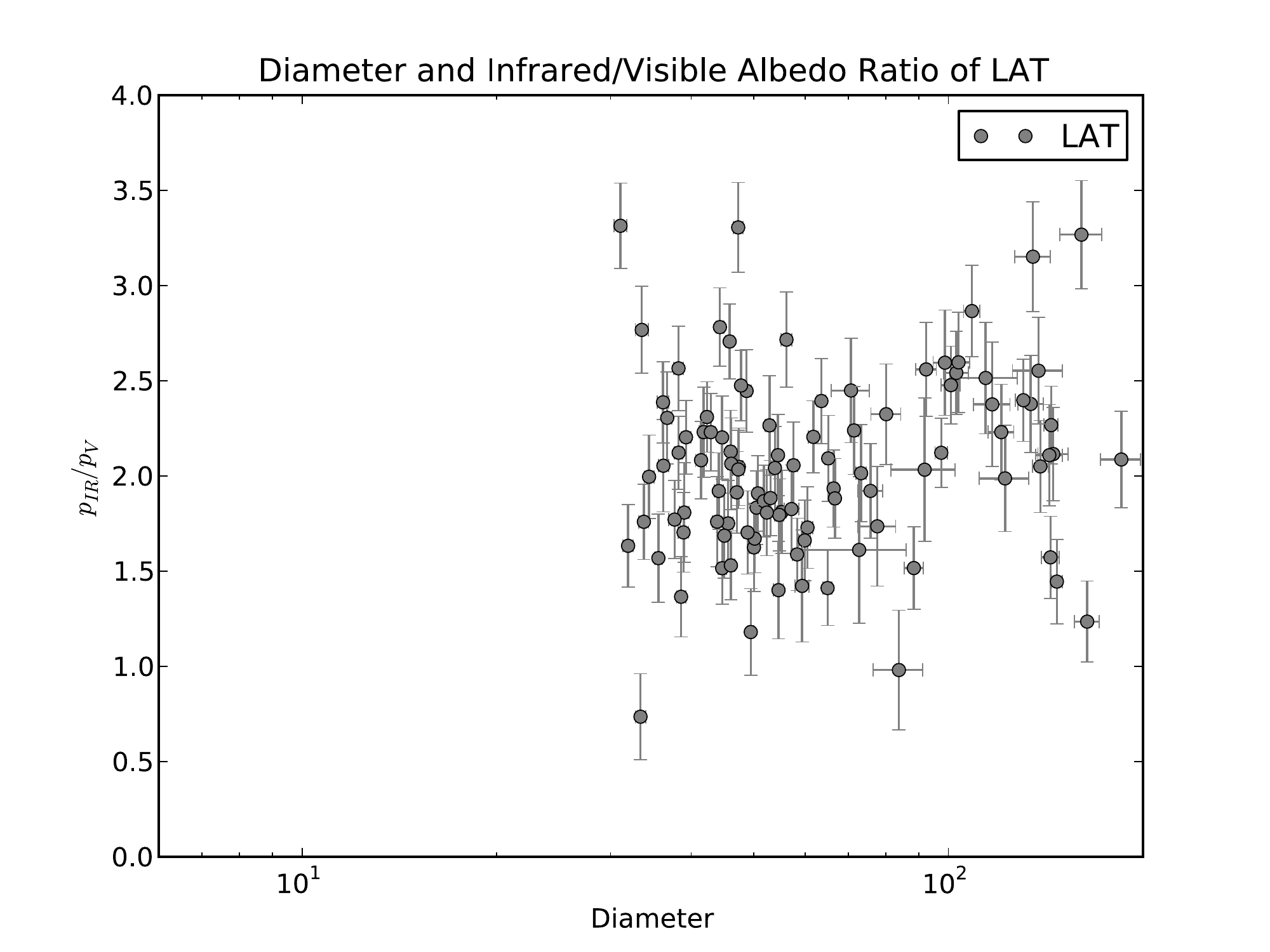}
\caption{The 100 long length-of-arc Trojans (LAT) objects for which diameter and infrared albedo were derived.}
\label{fig:DpIRfactor}
\end{center}
\end{figure}

In the LAT sample there are 100 objects (60 in the leading and 40 in the trailing) for which measurements are available in bands W1 and/or W2. As mentioned above this allows for the determination of the albedo at these wavelengths as these bands are almost exclusively reflected light. The results are shown versus diameter in Figure \ref{fig:DpIR}. While there appears to be a trend with higher $p_{IR}$ for smaller objects, this is highly likely to be due to observational biases as smaller objects with low infrared albedo most likely have fluxes below the sensitivity of WISE in W1 and/or W2.  Figure \ref{fig:DpIRfactor} shows the diameter versus ratio of $p_{IR}$ over $p_V$ for the 100 objects for which infrared albedo were fit. We see that larger objects are generally darker in the near-infrared than the small objects, but their slopes, i.e. ratio between $p_{IR}$ and $p_{V}$,  are steeper (i.e. redder).

\begin{figure}[t]
\begin{center}
\includegraphics[width=10cm]{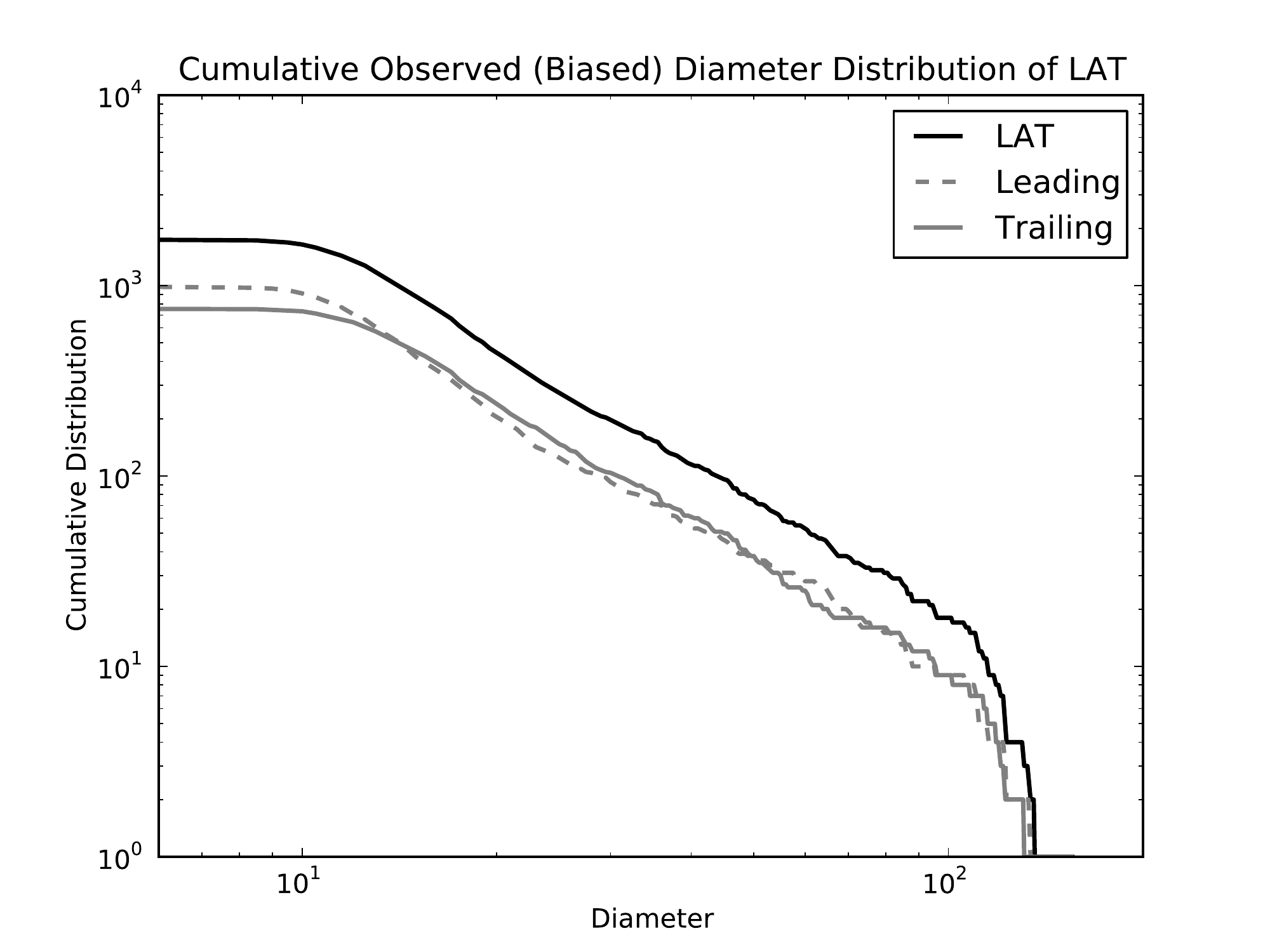}
\caption{The observed ({\bf biased}) cumulative size distribution of full sample (black), together with the leading (dashed grey) and trailing (solid grey) clouds. }
\label{fig:Dcum}
\end{center}
\end{figure}

From the LAT sample we can also look at the observed cumulative size distribution (see Figure \ref{fig:Dcum}). The observed distributions are remarkably similar, but we caution again that these are the raw distributions. As can be seen from Figure \ref{fig:ecliptic} the two clouds were not uniformly covered. The survey started partially into the leading cloud and the part closest to the planet was not observed until the Post-Cryogenic Mission, at which point only W1 and W2 were functioning, leading to a significant loss of objects in this part of the cloud. While the trailing cloud was completely covered during the cryogenic survey, the cloud's tail was in the galactic plane, close to the galactic center, at the time of observation. Careful debiasing is thus needed to properly compare the size distributions of the two clouds. 

In general the two clouds as observed are remarkably similar. Both the albedo and beaming distributions are the same to within the model errors, and even the infrared albedo distributions have similar means and widths. 

\section{Preliminary Debiasing of the Trojan Population}

In order to derive the true size and albedo distributions of the two Trojan clouds, we have to remove the inherent biases that exist in the WISE/NEOWISE sample. The strength of the NEOWISE survey is that it carried out a ``blind" search for moving objects, meaning that all moving objects were detected in the same fashion regardless of whether they were known beforehand. This uniformity allows the NEOWISE survey to be debiased independently of the biases of other surveys. Presented below is a preliminary attempt at debiasing, with focus on arriving an estimate on the relative abundance of objects within the two clouds. The full debiasing of this population will be presented in a future paper. 

\begin{figure}[t]
\begin{center}
\includegraphics[width=10cm]{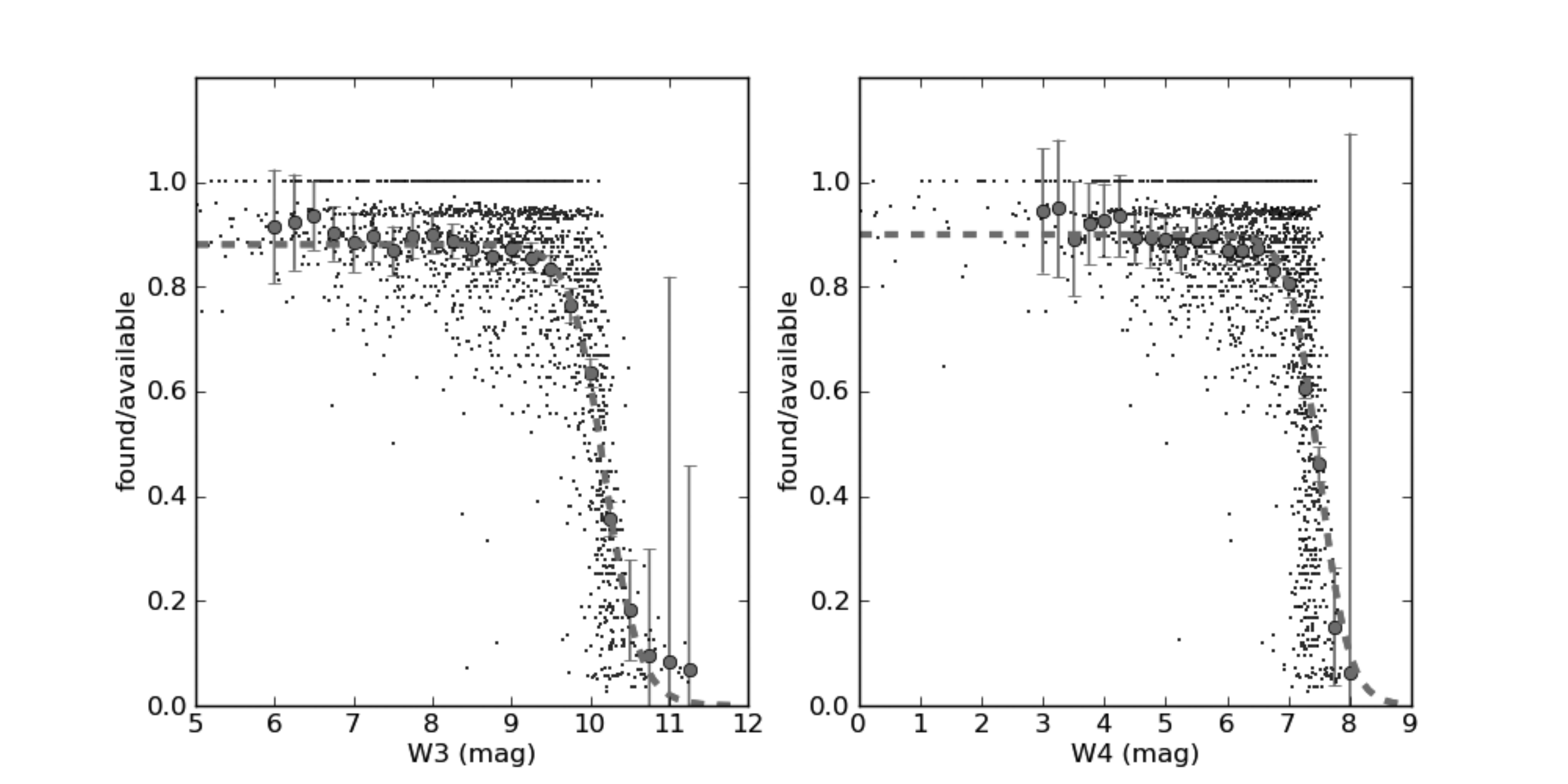}
\caption{The detection efficiency of the W3 and W4 bands as a function of magnitude as found using the KSSOPAL. }
\label{fig:debias_deteff}
\end{center}
\end{figure}

To model the NEOWISE survey bias, a high fidelity simulation of it was created. The time of observation, coordinates, orientation and footprint (47 x 47 arcminutes)  of all 1.2 million pointings used by WMOPS during the cryogenic survey was used to recreate the survey history. The WISE Known Solar System Object Possible Association List (KSSOPAL) was used to assess the survey detection efficiency at various locations across the sky. KSSOPAL used a list of known minor planet ephemerides to predict where the object should have been in each WISE frame and generated a list of probable matches. However, unlike WMOPS, it made no attempt at eliminating matches to inertially fixed sources such as stars and galaxies, nor did it check for spurious associations with artifacts or cosmic rays. We limited ourselves to the numbered asteroids in KSSOPAL as these in general have well-determined orbits. In order to reduce the possibility of spurious associations with stars or galaxies, we checked each source location from KSSOPAL against the WISE level 3 Atlas source table and used the $n$ out of $m$ statistics provided to search for sources that repeated; these sources were flagged. For each magnitude bin, the total number of available detections predicted by KSSOPAL was counted and compared to the total number of matches found. The estimate of single image completeness as function of flux for a particular region of the sky for bands W3 and W4 is shown in Figure \ref{fig:debias_deteff}. This completeness curve was computed for a number of different locations throughout the sky to sample the surveys sensitivity as a function of ecliptic and galactic latitude and longitude. The result is the probability that a moving object of a particular flux was detected by the WISE pipeline, and the detection probability curves P were fitted for both W3 and W4 using the following function:
\beq
    P = \frac{a_0}{2} (1 - \tanh(a_2 M - a_1)) + a_3
\eeq
where $M$ is the W3 or W4 magnitude and $a_i$ are the fitted coefficients. 

\begin{figure}[t]
\begin{center}
\includegraphics[width=10cm]{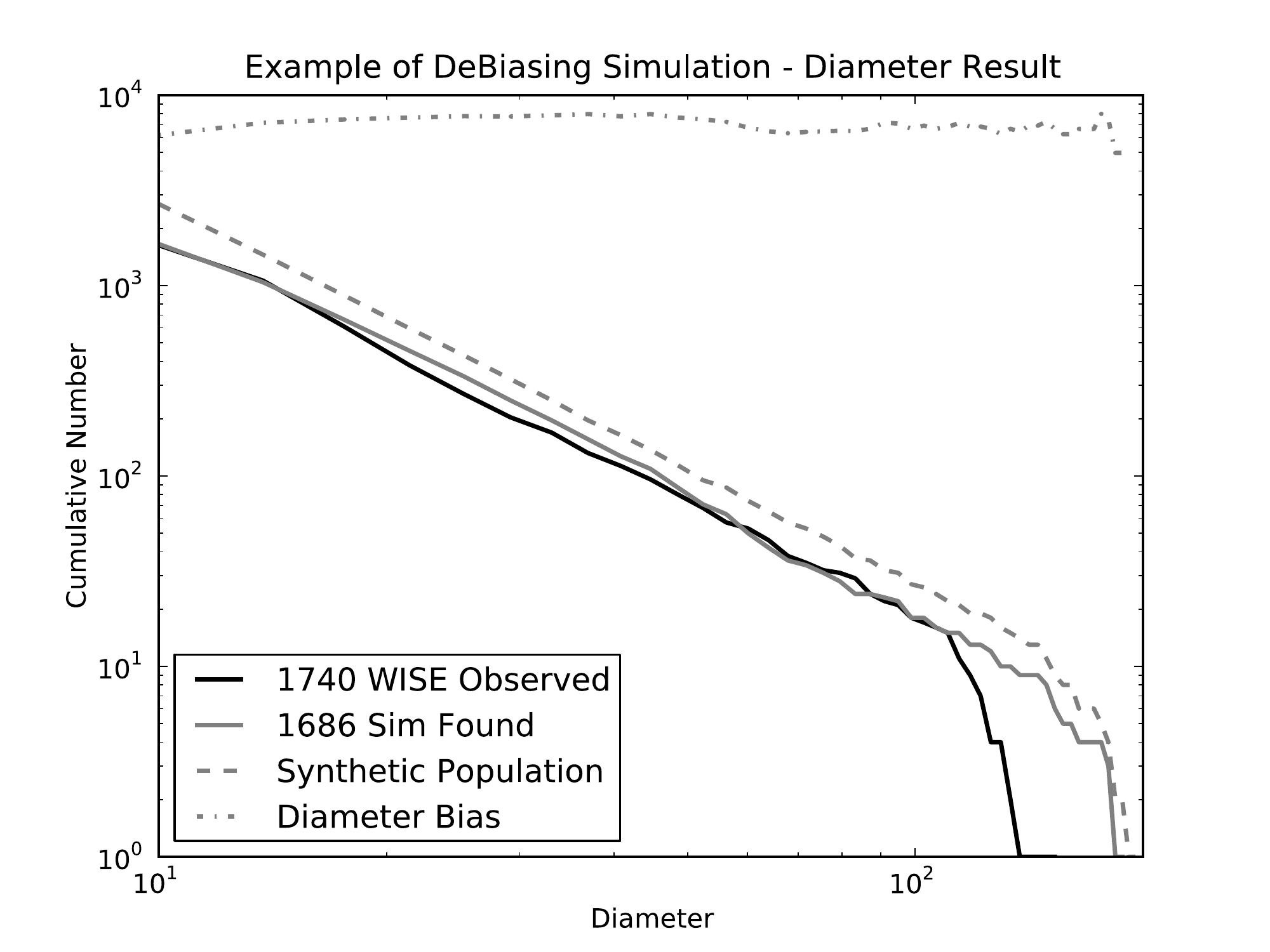}
\caption{Shown here is an example results from our debiasing simulations. The dashed grey line shows the synthetic population, consisting of 2750 objects with $ D > 10$km. The solid grey line gives the resulting simulated observed population. The black line gives the population with $D>10$km observed by WISE/NEOWISE. The dot-dashed grey line  shows the simulated population divided by the synthetic population, i.e. the bias introduced by the survey as a function of diameter, normalized by $10^4$. The plot only shows objects larger than 10km and observed arcs longer than 18 days.}
\label{fig:dia_debiasing}
\end{center}
\end{figure}

The orbital parameters for our synthetic population were created using the methodology described in \citet{Grav.2011a}, while the physical parameters were constructed using the observed distributions presented above. We assume here that the size, albedo and beaming distributions are the same for the two clouds. Looking at the results in the previous section this seems to be a reasonable assumption, but we will investigate this in deeper detail in future work. The synthetic populations were given a Gaussian albedo distribution with mean of $0.07$ and a fwhm of $0.03$. For the beaming a Gaussian with mean $0.88$ and a fwhm of $0.13$ were used. For the synthetic cumulative diameter distribution we use a power-law of the form:
\beq
	N(>D) = a_o D^{-\alpha}
\eeq
where we found $\alpha = 2$ to work very well for our preliminary debiasing \citep[c.f.][]{Jewitt.2000a}. 
In the following preliminary foray into debiasing of our survey we limited ourselves to objects with sizes larger than 10km, which yielded a sample of 1660 Jovian Trojans detected by NEOWISE/WISE. Figure \ref{fig:dia_debiasing} shows the comparison of the simulated and observed size distribution, which are in fair agreement. The difference between the synthetic population and the simulated survey is a measure of the bias introduced by the survey. 

\begin{figure}[t]
\begin{center}
\includegraphics[width=10cm]{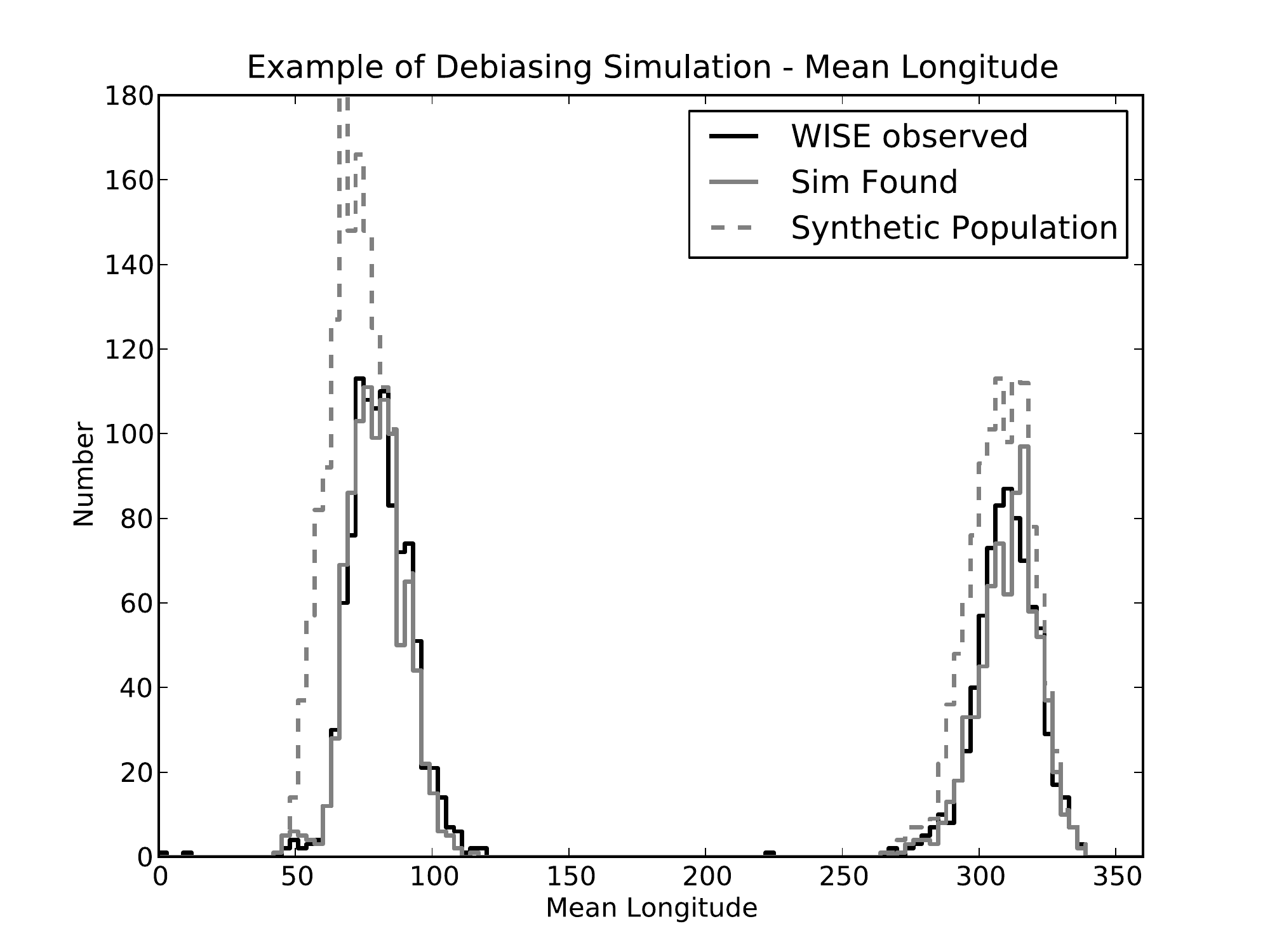}
\caption{Shown here is an example mean longitude results from our debiasing simulations. The dashed grey line gives the synthetic population used in the simulation. The grey solid line gives the resulting simulated population. The black solid  line gives the population observed by WISE/NEOWISE. The plots shows objects larger than 10km and observed arcs longer than 18 days.}
\label{fig:mlong_debiasing}
\end{center}
\end{figure}

Figure \ref{fig:mlong_debiasing} shows the relative number of objects in the two clouds using the preliminary debiasing simulations. We again caution that this is a preliminary result, but in this early work we were unable to derive any synthetic population with the equal number of objects in the two clouds that yielded simulated physical and dynamical distributions that were similar to that of the observed sample. Even adding in the SAT sample is unable to account for this relative number difference. If we assume that all the 141 objects in the trailing SAT sample and none of the 208 objects in the leading SAT sample are indeed Jovian Trojans (something that is highly unlikely), the number of trailing objects increases to 911. This highly unlikely scenario would only reduce the relative fraction from $\sim 1.4$ for the LAT sample alone to $\sim 1.2$. The lack of inclusion of the SAT sample in the preliminary debiasing is most likely the dominant error in determining the fraction of objects in the two clouds at this point. The two clouds thus have a fractional number of $N(\mbox{leading}) / N(\mbox{trailing}) \sim 1.4\pm0.2$, which is lower than the fraction of $1.6\pm0.1$ derived by \citet{Szabo.2007a} based on optical observations from the SDSS. This example shows, however, that full debiasing is the key to fully understand the similarities and differences between the two populations. This work is underway and will be presented in a future paper.

\subsection{Conclusions}

We have derived thermal models of 1739 Jovian Trojans together with a sample of 349 objects with observational characteristics that make them possible Trojans. This sample represents an increase by more than one order of magnitude in the number of Jovian Trojans with thermal measurements compared to previous surveys \citep{Tedesco.1992a,Tedesco.2002a,Fernandez.2003a,Fernandez.2009a,Ryan.2010a}. 

We find that the Jovian Trojan population is very homogenous for sizes larger than $\sim 10$km (close to the lower size limit for which  WISE is sensitive to these objects). The observed sample consists almost exclusively of low albedo objects, with the observed sample having a mean albedo value of $0.07\pm0.0.3$. The uniformly low albedos strengthens the notion that the population consists almost exclusively of C-, P- and D-type asteroids \citep{Gradie.1989a}. The beaming parameter was also derived for a large fraction of the observed sample, and is also very homogenous with an observed mean value of $0.88\pm0.13$. 

Preliminary debiasing of the survey shows our observed sample is consistent with the leading cloud containing more objects than the trailing cloud. We estimate the fraction to be $N(\mbox{leading})/N(\mbox{trailing}) \sim 1.4\pm0.2$, somewhat lower than with the $1.6\pm0.1$ value derived by \citet{Szabo.2007a}.  The size distribution is also found to broadly be consistent with the power-law slope found in \citet{Jewitt.2000a}, and work is underway to fully debias this interesting population of objects. 

\section{Acknowledgments}
This publication makes use of data products from the {\it Wide-field Infrared Survey Explorer}, which is a joint project of the University of California, Los Angeles, and the Jet Propulsion Laboratory/California Institute of Technology, funded by the National Aeronautics and Space Administration. This publication also makes use of data products from NEOWISE, which is a project of the Jet Propulsion Laboratory/California Institute of Technology, funded by the Planetary Science Division of the National Aeronautics and Space Administration. We gratefully acknowledge the extraordinary services specific to NEOWISE contributed by the International Astronomical Union's Minor Planet Center, operated by the Harvard-Smithsonian Center for Astrophysics , and the Central Bureau for Astronomical Telegrams, operated by Harvard University. We also thank the worldwide community of dedicated amateur and professional astronomers devoted to minor planet follow-up observations. This research has made use the NASA/IPAC Infrared Science Archive, which is operated by the Jet Propulsion Laboratory/California Institute of  Technology, under contract with the National Aeronautics and Space Administration.

\end{document}